\documentclass{aa}  

\usepackage{graphicx}
\usepackage{txfonts}
\usepackage{natbib}
\begin{document}

   \title{The density profile of Milky Way dark matter halo constrained from the OGLE microlensing sky map}

    \author{S. Lin
          \inst{1},
          W. Luo\inst{2,1}, 
          Y.F. Cai\inst{1,2},
          Q. Guo\inst{1},
          L. Wei\inst{3},
          B. Wang\inst{1},
          Q. Li\inst{4},
          C. Su\inst{1}
          \and
          A. Rodriguez\inst{5}
          }
   
   \institute{
   Department of Astronomy, School of Physical Science, University of Science and Technology of China, Hefei, Anhui 230026, China
   \and Institute of Deep Space Sciences, Deep Space Exploration Laboratory, Hefei, 230026, China 
   \and School of Physics and Astronomy, Sun Yat-sen University, 2 Daxue Road, Zhuhai 519082, China
   \and Department of Physics and Astronomy, University of Utah, Salt Lake City, UT 84102, USA
   \and Department of Astronomy, University of Michigan, 1085 S. University, Ann Arbor, MI 48109
    \email{all2b9s@mail.ustc.edu.cn}\\
    \email{wtluo@ustc.edu.cn}}

\titlerunning{The density profile of Milky Way dark matter halo constrained from the OGLE microlensing sky map}
\authorrunning{S. Lin et al}
   \date{Received September 29, 2024; accepted November 21, 2024}

 
  \abstract{
  We report the detection of a 282 $^{+34}_{-31}$ pc-sized core in the center of Milky Way dark matter halo at $68\%$ confidence level by using the micro-lensing event rate sky map data from the  Optical Gravitational Lensing Experiment (OGLE) survey. We apply the spacial information of the microlensing sky map and model it with the detailed Milky Way dark matter halo Core/Cusp profile, and the fraction of dark matter in the form of Mini Dark Matter Structure (MDMS, $f_{\rm MDMS}=\Omega_{\rm MDMS}/\Omega_{\rm DM}$), e.g. primordial black hole, earth-mass subhalos, floating planets and so on. 
 We find that this sky map can constrain both $f_{\rm MDMS}$ and the core size simultaneously without strong degeneracy while fully considering mass function of Milky Way stellar components from both the bulge and disk.}


\maketitle

%

\section{Introduction}

 \label{sec:intro}

The concordance cosmology is such a successful model
that it fits most of the observations with only half a dozen  parameters, e.g. Cosmic Microwave Background (CMB) \cite{Planck2020A&A}, Supernova Ia (SNIa)  \cite{Riess1998ApJ},  time delay projects based on strongly lensed Active Galactic Nuclei (AGNs) \cite{Birre2020A&A}, and weak lensing statistics \cite{Hikage_2019PASJ}. However, despite all these successes, tensions emerge among different cosmological probes, e.g. the recent Hubble tension between the Planck observation and SNIa \cite{Verde2019NatAs}, the ``lensing is low” issue on cosmological parameters between weak-lensing measurements and CMB \cite{Leauthaud_2017MNRAS}.

At small scales, there exist a series of “crises” as well, including the missing satellite problem (MSP), too big to fail (TBTF), as well as the core-cusp problem (CCP)  \cite{deBlok2010AdAst}. Especially, the core-cusp issue, where the inferred dark matter core structure from nearby dwarf galaxies contradicts the Navarro-Frenk-White (NFW) profile, a cuspy profile from pure cold dark matter simulations \cite{NFW_1997ApJ}, leads to the extensive study on the formation mechanism of the core structure either from an exotic self interacting dark matter (SIDM) model \cite{Spergel2000PhRvL, JiangF2023MNRAS.521.4630J}, or an astrophysical model from baryonic feedback and merging, dynamical friction, e.t.c. [see review papers of \cite{Salucci2019A&ARv,Popolo2022arXiv}]. 
Moreover, the enigmatic Gould Belt can be created by a “dark matter core" colliding with the gas disk of Milky Way \cite{Diemand2008Natrue,Bekki2009MNRAS}. 
However, so far there is no conclusive observational evidence to either confirm or reject the aforementioned core structure of Milky Way. This is largely due to the fact that dark matter itself can not be directly observed and the rotation curve of Milky Way is not as informative as nearby dwarf {galaxies \cite{Gentile2004MNRAS, Bekki2009MNRAS, Salucci2017MNRAS}.
New probes are then necessary for mapping the dark side of the Milky Way and comparing them to simulations, which will boost our understanding of the core formation of Milky Way halo and the like.

The method we develop here is based on the assumption that a fraction of dark matter are in the form of primordial black holes \cite{Starobinskii1979A&A,Bernard2020ARNPS,Misao2018CQGra..} , dark matter halos with hundreds of earth masses \cite{Wang2020Natur}, or to a very recent work \cite{Delos2023mnras}: the so-called dark matter minihalos with extremely high internal density of $10^{12}M_{\odot}/\rm pc^{3}$.  We generalize it as mini dark matter structures (MDMS) which follow the density profile of Milky Way dark matter halo and can induce the micro-lensing event when intervening the line between the background star and the observer.

\begin{figure*}[ht]
\includegraphics[width=18cm,height=5cm]{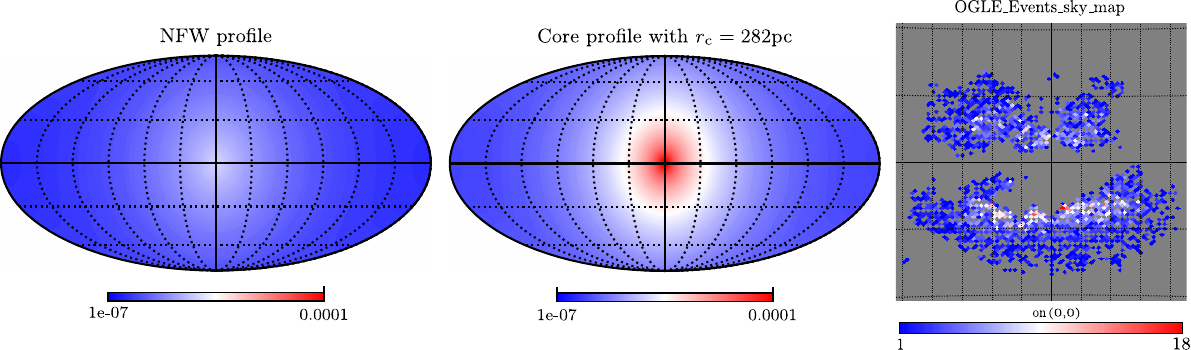}
    \caption{The sky-maps are plotted in Mollweide projection.
    Left panel: Micro-lensing event rate given an NFW density profile; 
    Middle panel: 282pc core size event rate; 
    Right panel: The sky map of 5790 OGLE micro-lensing events. The difference between graticules are 2 degrees in longitude and 4.5 degrees in latitude.
    As we are unable to reach all the detailed data of micro-lensing events used to build the catalog, we can not plot an event-rate sky map [Readers can find it in Fig.24 of \cite{Mroz2020ApJS}]. }
\label{fig:ogle}
\end{figure*}
Fig.\ref{fig:ogle} presents the patterns of micro-lensing event rate map between the cuspy NFW profile (left panel) and the cored density profile (middle panel).
The right panel is the event rate sky map of OGLE newest data release \cite{Mroz2019APJS}. By fitting the OGLE sky map, we can constrain the dark matter density profile.

Here, we serendipitously found that the spatial distribution of micro-lensing event rate sky map targeting the Galactic center from Optical Gravitational Lensing Experiment (OGLE) \cite{Mroz2020ApJS} can put strong constraints simultaneously on both Mini Dark Matter Structures (MDMS) fraction, and the core size of the central part of Milky Way dark matter halo. The inferred size of the dark core is 282 $^{+34}_{-31}$pc at 68\% confidence level. Further, this result can put constraints on both dark matter models beyond cold dark matter (CDM) or the baryonic physics of the Milky Way \cite{wetzel2023ApJS}. Either way, this discovery opens another window to peek at the mystery of the core-cusp problem.

\section{Microlensing model}
Our model extends the micro-lensing geometry and notations in Niikura et al. 2019 \cite{Niikura2019PRD} by fully employing the spacial information.
We take the x-direction to be along the line connecting the Galactic center and the Earth (the observer's position),   
with the assumption that the Earth is located at $(x, y, z) = (8 \mathrm{kpc}, 0, 0)$. 
And the y-direction is the rotating direction of the Earth in the Galactic disk plane together with the z-direction being perpendicular to the plane. 

In this paper, we denote the mass of the lens as $M$, while the source-observer, lens-observer, and source-lens distance as $d_{\mathrm{s}}$, $d_{\mathrm{l}}$ and $d_{\mathrm{ls}}=d_{\mathrm{s}}-d_{\mathrm{l}}$.
In our coordinate system, a lens at $d_{\mathrm{l}}$ will be located at: $x = 8\mathrm{kpc}-d_{\mathrm{l}}\cos l\cos b$, $y = d_{\mathrm{l}}\cos b \sin l$, $z =d_{\mathrm{l}} \sin b$.
As the OGLE survey mainly focuses on the source near the Galactic center, we assume $d_{\mathrm{s}} = 8$ kpc for all the sources as have been done in Niikura et al. 2019 \cite{Niikura2019PRD}.
With the above notation, the Einstein radius $R_{\mathrm{E}}$ for a given mass $M$ can be expressed as:

\begin{equation}
    R_\mathrm{E} = \sqrt{\frac{4GM}{c^2}\frac{d_\mathrm{l} d_{\mathrm{ls}}}{d_\mathrm{s}}}
    \label{eq:Re}
\end{equation}

For simplicity, we consider that all the lenses have an identical mass of $M$ for MDMS, but stellar mass function (see the following section ) for bulge and disk contribution. Both of them follow the Maxwell-Boltzmann velocity distribution (we also test a different velocity distribution to examine if the core size varies accordingly in the supplemental appendix). To be specific, as we are only interested in the relative velocity between lenses and source that is perpendicular to the $x$ direction, the velocity distribution has the following form:
\begin{equation}
    f(\textbf{v}_\bot) = \frac{1}{2\pi \sigma_y \sigma_z}                       \exp\left[{\frac{(v_\bot \cos\theta-\bar{v}_y)^2}{\sigma_y^2}+\frac{(v_\bot \sin\theta-\bar{v}_z)^2}{\sigma_z^2}}\right]
\end{equation}
where $\theta$ represents the direction of $\textbf{v}_\bot$, $\sigma_y$, $\sigma_z$ are the velocity dispersion of lens in the $y$ and $z$ directions and $\bar{v}_y$, $\bar{v}_z$ are the mean velocity along $y$ and $z$ directions.

With the above assumptions, the micro-lensing event rate for timescale $t_\mathrm{E}$ along a certain direction is then:
\begin{align}
\nonumber
    \frac{\mathrm{d} \Gamma_{\mathrm{d}}}{\mathrm{d} t_{\mathrm{E}}}
    &=\pi
    \int \mathrm{d} (\ln M) \frac{\mathrm{d} n_{\mathrm{lens}}(M)}{\mathrm{d} \ln M}
    \int_{0}^{d_{\mathrm{s}}}\mathrm{d}d_{1}\frac{ \rho_{\mathrm{lens}}\left(d_{1}\right)}{M}\\
    &\int_{-\pi / 2}^{\pi/2}\mathrm{~d\theta}v_{\perp}^{4}f_{\mathrm{lens}}\left(v_{\perp},\theta\right)
\end{align}
where $t_\mathrm{E}$ is the microlensing event time scale given by $t_\mathrm{E} =(2R_{\mathrm{E}} \cos\theta)/(v_\bot)$. 
And $\rho_{\mathrm{lens}}\left(d_{1}\right)$ is the density profile for lenses' distribution along the line of sight at the distance of $d_{\mathrm{l}}$, whose specific expression depends on its kind, i.e. dark matter, bulge stellar and disk stellar. The $\rho_{lens}$ of MDMS can be simply expressed as $f_{MDMS}\rho_{halo}$, and $f_{MDMS}=\frac{\Omega_{MDMS}}{\Omega_{DM}}$ denoting the fraction of MDMS as dark matter.  The term in the right side of Eq.3  $\frac{\mathrm{d} n_{\mathrm{lens}}(M)}{\mathrm{d} \ln M}$ is the mass function of lens. We assume a delta mass function for MDMS and will discuss the detailed mass function of stellar components in the following sections.

As the density profile is only considered in the integration along the line of sight, the angular distribution of the density profile is encoded inside the event rate distribution.

\subsection{Dark matter profile}
For cusp-like dark matter halo, we adopt the NFW density profile:
\begin{equation}
    \rho_{\mathrm{NFW}}(r) = \frac{\rho_{\mathrm{c}}}{(r/r_{\mathrm{s}})(1+r/r_{\mathrm{s}})^2}
\end{equation}
where $r$ is the distance to the Galactic center, $\rho_\mathrm{c} = 4.88 \times 10^6 M_\odot/\mathrm{kpc}^3$ and $r_\mathrm{s} = 21.5\mathrm{kpc}$. For a core-liked halo, we choose the Burkert density profile \cite{Burkert1995APJ, salucci2000ApJ}:
\begin{equation}
    \rho_{\mathrm{Bk}}(r) = \frac{\rho_{\mathrm{b}}}{(1+r/r_{\mathrm{b}})(1+(r/r_{\mathrm{b}})^2)}
\end{equation}
where $r_{\mathrm{b}}$ stands for the core size and $\rho_{\rm b}$ is determined by the halo mass. Here we assume the Milky Way halo mass to be of $10^{12} M_\odot$ \cite{Klypin_2002}. 
The density profile and velocity profile we use for each component of lens are shown in Table.1
\cite{Niikura2019PRD}.

\begin{table*}[ht]
\centering
\begin{tabular}{|c|l|l|l|} 

 \hline 
 Lens' kind
 & Density profile $\rho[M_\odot/ \mathrm{pc}^3]$
 & Velocity profile($\mu,\sigma$) [km/s]
 \\ 
 \hline 
 
Bulge 
& $1.04\times10^6(\frac{s}{0.482\mathrm{pc}})^{-1.85}$, $(s<938\mathrm{pc})$ 
& $f_y : \{-220(1-\alpha),100\sqrt{1+\alpha^2}\}$ 
\\
& $3.53$ $K_0(\frac{s}{667 \mathrm{pc}})$, ($s\ge938\mathrm{pc}$)
& $f_z : \{0,100\sqrt{1+\alpha^2} \}$
\\
\hline
Disk
&  $0.06\times\mathrm{exp}[-\{\frac{R-8000}{3500}+\frac{z}{325}\}]$
& $f_y : \{220\alpha, \sqrt{(\kappa\delta+30)^2+(100\alpha)^2}\}$
\\&
& $f_z : \{0, \sqrt{(\lambda\delta+30)^2+(100\alpha)^2}\}$
\\
\hline
MDMS
&  Cusp: $4.88\times10^{-3}\frac{1}{(r/r_\mathrm{s})(1+r/r_\mathrm{s})^2}$
& $f_y : \{-220(1-\alpha),\sqrt{220^2+(100\alpha)^2}\}$
\\
& Core: $\frac{\rho_{\mathrm{b}}}{(1+r/r_{\mathrm{b}})(1+(r/r_{\mathrm{b}})^2)}$
& $f_z: \{0,\sqrt{220^2+(100\alpha)^2}\}$
\\
 
 \hline

 \end{tabular}
   \caption{Density profile and velocity profile for each kind of lens, where $\alpha = d_\mathrm{l}/d_\mathrm{s}$, 
    $\kappa = 5.625\times10^{-3}\mathrm{km/s/pc}$, $\lambda = 3.75\times10^{-3}\mathrm{km/s/pc}$, $\delta = (8000-x)\mathrm{pc}$,
    $K_0(x)$is modified Bessel function and R=$(x^2+y^2)^\frac{1}{2}$, $s^4=R^4+(z/0.61)^4$, $r^2=x^2+y^2+z^2$
    }
  \label{density profile}
\end{table*}

\subsection{Mass function of stellar components}
In order to estimate the event rate of stellar components in bulge and disk, we need to get the mass function of each kind. 
For this purpose, we assume the Kroupa broken power-law initial mass function (IMF)\cite{Kroupa2000MNRAS}.

\begin{align}
\frac{\mathrm{d} n_{\mathrm{s}}(M)}{\mathrm{d} \ln M}
& = \begin{cases}
A_{\mathrm{MS}}\left(\frac{M}{M_\mathrm{break}}\right)^{1-\alpha_{\mathrm{MS} 1}} & \left(M \leq M_\mathrm{break}\right) \\
A_{\mathrm{MS}}\left(\frac{M}{M_\mathrm{break}}\right)^{1-\alpha_{\mathrm{MS} 2}} & \left(M\geq M_\mathrm{break}\right)\end{cases}
\end{align}

Following previous work \cite{Niikura2019PRD}, we assume all stars within the initial mass range $[1 \leq  M_\mathrm{break}/M_\odot \leq 8]$ evolves into White Dwarfs following initial-final mass relation of :$M_{\mathrm{WD}} = 0.339+0.129M_{\mathrm{init}}$ and stars with $[8 \leq  M_\mathrm{break}/M_\odot \leq 20]$ into neutron stars following a Gaussian distribution with mean value $M_{\mathrm{final}} = 1.39 M_\odot$ and width $\sigma = 0.12 M_\odot$.

We choose the newer data from Mr\'{o}z et al. 2020 \cite{Mroz2020ApJS} instead of the previous version based on Mr\'{o}z et al. 2017 \cite{Mroz2017Nature} which only contains 9 fields.
Here, we set $\alpha_{\mathrm{MS2}} = 2, M_\mathrm{break} = 0.5 M_\odot$, and leave $\alpha_\mathrm{MS1}$ as free parameter to be sampled by MCMC. We choose $\alpha_\mathrm{MS1} = 1.1$ as fiducial value, which is consistent with result from Sagittarius Window Eclipsing Extrasolar Planet Search (SWEEPS)\cite{Calamida2015APJ}. 

\subsection{Fitting to data}
Before proceeding with MCMC, we need to subtract the contribution of bulge and disk components in event rates sky map. That is:
\begin{equation}
    \Gamma_{\mathrm{fixed,k}} = \Gamma_{\mathrm{OGLE,k}} - \Gamma_{\mathrm{bulge,k}} - \Gamma_{\mathrm{disk,k}}
\end{equation}
where $\rm k$ is the index of fields in OGLE survey \cite{Mroz2020ApJS}. Afterwards, we neglect all the fields with negative values and use the remaining 55 fields, 
where we choose to fit two variables: the fraction of MDMS ($f_{\mathrm{MDMS}}$) and the size of the core ($\log_{10} r_{\mathrm{b}}$) by fitting the theoretical event rate $\Gamma_{\mathrm{Core}}$ to the event rate at the directions of 55 fields. For simplicity, we assume a log-normal likelihood function:
\begin{equation}
    \ln(p) = \sum\limits^{55}_{\mathrm{k}=1} -\frac{(\ln(\Gamma_{\mathrm{Core,k}})-\ln(\Gamma_{\mathrm{fixed,k}}))^2}
    {2\sigma^2_{\mathrm{k}}}+ \ln(\sigma_\mathrm{k})
\end{equation}
$\sigma_\mathrm{k}$ in the above formula is given by 
$\sigma_\mathrm{k}^2 = \ln(1+\sigma^2_{\mathrm{OGLE,k}}/\Gamma^2_{\mathrm{fixed,k}}) $ in which $\sigma_{\mathrm{OGLE,k}}$ is the error of event rate from the OGLE data.

Based on the likelihood function, we use the python package \texttt{EMCEE}\cite{EMCEE2013} to run MCMC with 20 walkers and 3500 steps each after burn-in processes of 500 steps. The posteriors of the two parameters are therefore sampled from the chains. 

\section{Results}
\begin{figure}
\includegraphics[width=0.95\columnwidth]{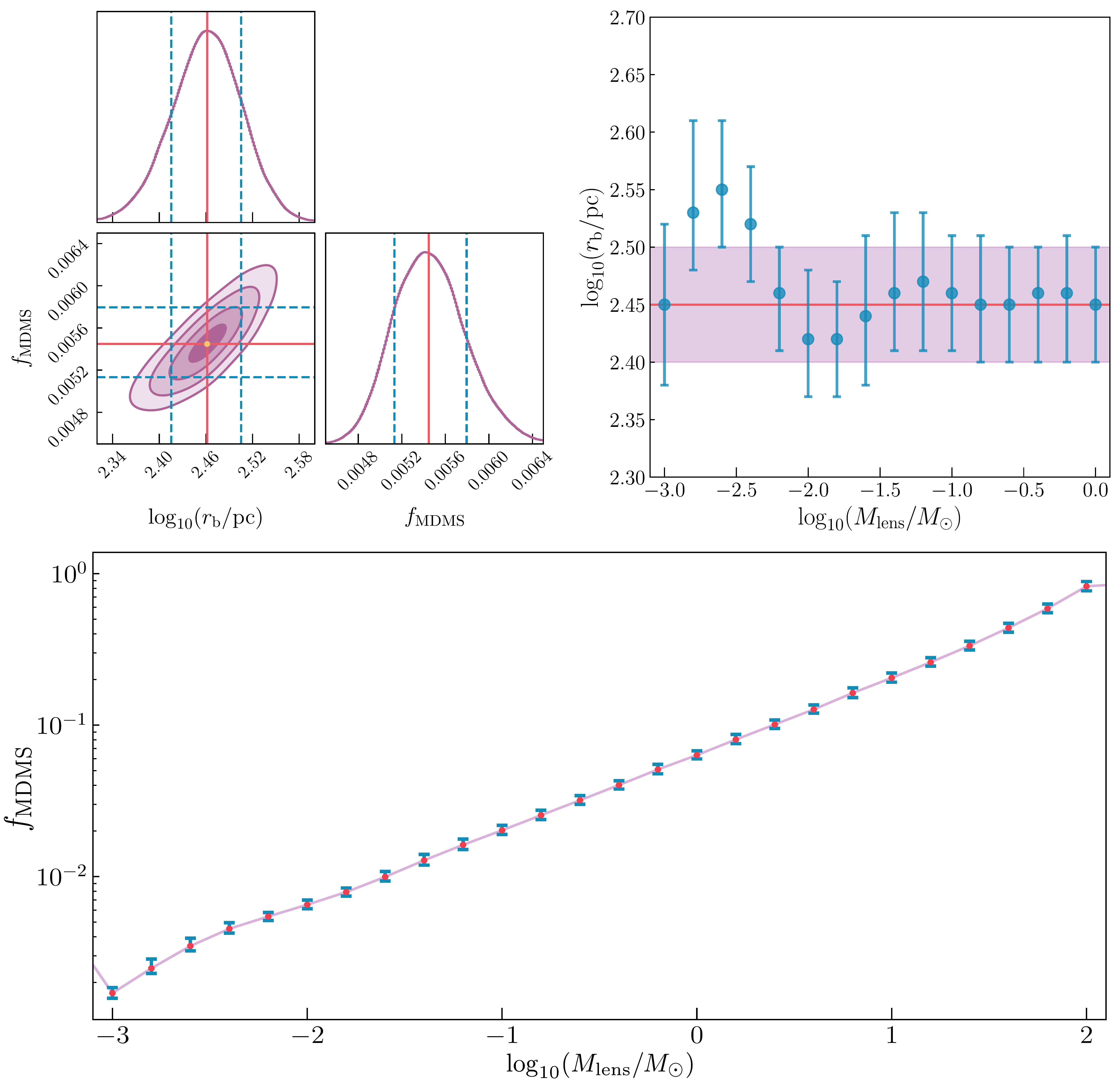}
    \caption{ 
    \textit{Top left: } The corner plot of the two parameters $f_{\mathrm{MDMS}}$ and core size $r_b$ with lens mass of $10^{-2.2}M_\odot$. The red solid lines are the median value of the posterior distribution and the blue dashed lines denote the 68\% confidence interval. The yellow point is the median value of MCMC sampled posterior and different contours in the corner plot denote the different $\sigma$ levels. \\
    \textit{Top right: } The core size as a function of lenses' mass. The red solid line is the median value from the left panel for $r_b$ with one sigma range as the purple belt. The different core sizes sampled from different lens mass is shown as the data points with error bars. The value remains almost unchanged within one sigma level.\\
    \textit{Bottom: } Fraction of MDMS $f_{\mathrm{MDMS}}$ from MCMC as function of lenses' mass. The red points are the median with blue errorbars showing the one sigma range for each mass value. The purple solid line connects the median values.}
    \label{fig:mcmc}
\end{figure}
We perform a test by calculating the Bayesian factor for both the cuspy profile (NFW formula) and the core profile (Burkert formula) with the nested sampling Monte Carlo algorithm
MLFriends \cite{NestSam2016, NestSam2019} using the UltraNest package \cite{Ultranest2021}
. For all the lens mass we tested, the ratios between Bayesian factors of the Burkert profile and NFW profile are larger than $10^{3}$. This illustrates that the Burkert density profile is much more probable than the NFW profile. 

The major results are shown in Fig.2, the top left panel presents the posterior distribution of core size ($\rm log_{10}(r_b/pc)$) and the fraction of MDMS ($f_{MDMS}$) assuming the lens mass of $\sim 10^{-2}M_{\odot}$. The latter varies significantly as a function of lens mass due to the OGLE survey cadence (the bottom panel of Fig.2). Nonetheless, the core size remains consistent with the fiducial value that the lens is set to be one solar mass (red solid line with one sigma pink shaded region in the top right panel of Fig.2). The effective range is between $10^{-3} M_\odot$ and $10^{0} M_\odot$, where the constraint is the most effective based on OGLE survey data.

Astonishingly, for lens' mass around $10^{-3} M_\odot$, $f_{\mathrm{MDMS}}$ is about $10^{-3}$. That means, to the maximum, $10^9 M_\odot$ dark matter is in the form of MDMS, which is a much stronger constraint than previous result\citep{Niikura2019PRD}. The dependence of $f_{\mathrm{MDMS}}$ on lens mass is due to the cadence of OGLE survey which is in $[0.1\mathrm{days}, 300 \mathrm{days}]$. As a result, when lens' mass is too large($>1 M_\odot$) or too small($<10^{-3} M_\odot$), the constraint on $f_{\mathrm{MDMS}}$  is approaching to $1$. The bottom panel of Fig.2 clearly shows the dependence of  $f_{\mathrm{MDMS}}$ on lens stellar mass, which is purely introduced by the OGLE survey cadence.

\section{Discussion and summary}

In a review on the density profile of dark matter halos by Paulo Salucci et al 2019 \citep{Salucci2019A&ARv} concludes that the Milky Way dark matter halo has a core structure with the size between 1-100kpc. \cite{Nesti2013JCAP} estimated the core size of Milky Way dark matter halo to be about 10kpc using multiple dynamical data, such as MW terminal velocities in the region inside the solar circle, circular velocity as recently estimated from maser star forming regions at intermediate radii, and velocity dispersions of stellar halo tracers for the outermost Galactic region. The relation between our result and theirs is worth being probed if the smaller structure is an additional one besides the larger core.

In \cite{Chan2015MNRAS}, they generate a suite of simulations based on the same code of FIRE, focusing on the core size of galaxies with a wide range of stellar mass and feedback. Their results present much larger core sizes than ours ranging from 1.2kpc to 2.0kpc for Milky Way dark matter halo. Our results shows a much smaller core size that caused by the baryonic feedback in those simulations. 

As the first try to constrain the core of Milky Way dark matter halo using the OGLE micro-lensing sky map, we do realize the limitation of the OGLE data for our constraints, which is targeting the Milky Way bulge. Before we draw any solid conclusion of the implications for hydro simulations and dark matter particle properties (as described in the next paragraph) with the core size, we will further explore other means to test our results. In an another upcoming work, we select micro-lensing event candidates in Milky Way halo using a technique called "static micro-lensing" (\cite{Guo2024arXiv241102161G}, He et al in prep.). We will further explore the core size based on those new data. In future, powerful time domain surveys such as LSST \cite{Ivezic2019ApJ...873..111I}, WFST\citep{Wang2023SCPMA} will also provide valuable microlensing events in Milky Way halo.

On the other hand, phenomenological models beyond CDM provide another means to interpret the core structure formation, ranging from a solitonic core of ultra-light dark matter (ULDM) \cite{easther2020PASA}, or fuzzy dark matter (FDM) \cite{Burkert2020APJ}, to self-interacting dark matter (SIDM) \cite{SIDM2013MNRAS} as well as weakly interacting massive particles (WIMP) \cite{deSwart2017NatAs}. 
However, the core size of the dark matter halo behaves differently in the stellar feedback scenario from one those models, specifically SIDM model\cite{Rocha2013MNRAS}. The former shows the core size peaks around $\rm log_{10}(M_{\ast}/M_{\odot})\sim 10.0$  while the latter illustrates a monotonic increase as a function of halo mass. In general, different dark matter particle models can potentially be tested by astrophysical observations, such as modulated Einstein rings from multiple image systems of strong-lensing events \cite{Smoot2023NatAs}. We present a novel method to probe the dark matter density profile by using the micro-lensing sky map, which can be used to further constrain various mechanisms of core formation.

To summarize, we apply the OGLE micro-lensing sky map to obtain by far the tightest constraint on the core size of Milky Way dark matter halo. 
The core size value is $282 ^{+34}_{-31} \mathrm{pc}$ and is independent of $M_\mathrm{lens}$ within a wide mass range. This result can potentially put stronger constraints on the cross section of SIDM particles, the mass of ULDM/FDM. The core size can constrain the strength of star formation process of Milky Way.
We acknowledge that we do not consider an off-center between the dark matter halo potential center and the Galactic center in our modeling which can be another interesting issue to probe.  We also notice that the OGLE event rate sky map we use only locates at the Galactic center, and more data beyond the Galactic center region will greatly improve the constraint based on our model. 

In the future, providing a wider range survey of micro-lensing event rate map (potentially Gaia archive data\cite{Gaia2021A&A}), or novel statistical measures of micro-lensing event rate sky map, our method can be extended to a series of studies on the detailed structures of Milky Way dark matter halo.


We are grateful to Shude Mao, Ruizhi Yang, Weicheng Zang, Elisa Ferreira, and Christopher J. Miller for valuable discussions. 
This work is supported in part by NSFC (12192224, 12003029,  11833005),  National Key R\&D Program of China (2021YFC2203100), the Frontier Scientific Research Program of Deep Space Exploration Laboratory (No. 2022-QYKYJH-HXYF-012), by CAS young interdisciplinary innovation team (JCTD-2022-20), by 111 Project for "Observational and Theoretical Research on Dark Matter and Dark Energy" (B23042), by Fundamental Research Funds for Central Universities, by CSC Innovation Talent Funds, by USTC Fellowship for International Cooperation, by USTC Research Funds of the Double First-Class Initiative, by CAS project for young scientists in basic research (YSBR-006).

\paragraph{Data availability.} The code supporting this work is available from the corresponding authors upon reasonable request.

The \texttt{EMCEE} package is available under MIT License in:  https://emcee.readthedocs.io/en/stable/. 

The Ultranest package is available in https://johannesbuchner.github.io/UltraNest/.

The microlensing modeling code is available in
https://github.com/all2b9s/Static-microlensing .

The OGLE catalogue is publicly available in:  https://ogle.astrouw.edu.pl/.\\

\bibliographystyle{aa}
\bibliography{reference}

\begin{thebibliography}{51}
\expandafter\ifx\csname natexlab\endcsname\relax\def\natexlab#1{#1}\fi

\bibitem[{{Amruth} {et~al.}(2023){Amruth}, {Broadhurst}, {Lim}, {Oguri}, {Smoot}, {Diego}, {Leung}, {Emami}, {Li}, {Chiueh}, {Schive}, {Yeung}, \& {Li}}]{Smoot2023NatAs}
{Amruth}, A., {Broadhurst}, T., {Lim}, J., {et~al.} 2023, Nature Astronomy, 7, 736

\bibitem[{Bekki(2009)}]{Bekki2009MNRAS}
Bekki, K. 2009, Monthly Notices of the Royal Astronomical Society: Letters, 398, L36

\bibitem[{{Birrer} {et~al.}(2020){Birrer}, {Shajib}, {Galan}, {Millon}, {Treu}, {Agnello}, {Auger}, {Chen}, {Christensen}, {Collett}, {Courbin}, {Fassnacht}, {Koopmans}, {Marshall}, {Park}, {Rusu}, {Sluse}, {Spiniello}, {Suyu}, {Wagner-Carena}, {Wong}, {Barnab{\`e}}, {Bolton}, {Czoske}, {Ding}, {Frieman}, \& {Van de Vyvere}}]{Birre2020A&A}
{Birrer}, S., {Shajib}, A.~J., {Galan}, A., {et~al.} 2020, Astronomy \& Astrophysics, 643, A165

\bibitem[{Bozorgnia \& Bertone(2017)}]{velocity_distribution_dm}
Bozorgnia, N. \& Bertone, G. 2017, International Journal of Modern Physics A, 32, 1730016

\bibitem[{{Buchner}(2016)}]{NestSam2016}
{Buchner}, J. 2016, Statistics and Computing, 26, 383

\bibitem[{{Buchner}(2019)}]{NestSam2019}
{Buchner}, J. 2019, Publications of the Astronomical Society of the Pacific, 131, 108005

\bibitem[{{Buchner}(2021)}]{Ultranest2021}
{Buchner}, J. 2021, The Journal of Open Source Software, 6, 3001

\bibitem[{Burkert(1995)}]{Burkert1995APJ}
Burkert, A. 1995, The Astrophysical Journal, 447

\bibitem[{Burkert(2020)}]{Burkert2020APJ}
Burkert, A. 2020, The Astrophysical Journal, 904, 161

\bibitem[{Calamida {et~al.}(2015)Calamida, Sahu, Casertano, Anderson, Cassisi, Gennaro, Cignoni, Brown, Kains, Ferguson, Livio, Bond, Buonanno, Clarkson, Ferraro, Pietrinferni, Salaris, \& Valenti}]{Calamida2015APJ}
Calamida, A., Sahu, K.~C., Casertano, S., {et~al.} 2015, The Astrophysical Journal, 810, 8

\bibitem[{{Carr} \& {K{\"u}hnel}(2020)}]{Bernard2020ARNPS}
{Carr}, B. \& {K{\"u}hnel}, F. 2020, Annual Review of Nuclear and Particle Science, 70, 355

\bibitem[{{Chan} {et~al.}(2015){Chan}, {Kere{\v{s}}}, {O{\~n}orbe}, {Hopkins}, {Muratov}, {Faucher-Gigu{\`e}re}, \& {Quataert}}]{Chan2015MNRAS}
{Chan}, T.~K., {Kere{\v{s}}}, D., {O{\~n}orbe}, J., {et~al.} 2015, Monthly Notices of the Royal Astronomical Society, 454, 2981

\bibitem[{Croon {et~al.}(2020)Croon, McKeen, \& Raj}]{CroonPRD2020}
Croon, D., McKeen, D., \& Raj, N. 2020, Physical Review D, 101

\bibitem[{{de Blok}(2010)}]{deBlok2010AdAst}
{de Blok}, W.~J.~G. 2010, Advances in Astronomy, 2010, 789293

\bibitem[{{de Swart} {et~al.}(2017){de Swart}, {Bertone}, \& {van Dongen}}]{deSwart2017NatAs}
{de Swart}, J.~G., {Bertone}, G., \& {van Dongen}, J. 2017, Nature Astronomy, 1, 0059

\bibitem[{{Del Popolo} \& {Le Delliou}(2022)}]{Popolo2022arXiv}
{Del Popolo}, A. \& {Le Delliou}, M. 2022, arXiv e-prints, arXiv:2209.14151

\bibitem[{{Delos} \& {Silk}(2023)}]{Delos2023mnras}
{Delos}, M.~S. \& {Silk}, J. 2023, Monthly Notices of the Royal Astronomical Society: Letters, 520, 4370

\bibitem[{Diemand {et~al.}(2008)Diemand, Kuhlen, Madau, Zemp, Moore, Potter, \& Stadel}]{Diemand2008Natrue}
Diemand, J., Kuhlen, M., Madau, P., {et~al.} 2008, Nature, 454, 735

\bibitem[{{Foreman-Mackey} {et~al.}(2013){Foreman-Mackey}, {Hogg}, {Lang}, \& {Goodman}}]{EMCEE2013}
{Foreman-Mackey}, D., {Hogg}, D.~W., {Lang}, D., \& {Goodman}, J. 2013, Publications of the Astronomical Society of the Pacific, 125, 306

\bibitem[{{Gaia Collaboration} {et~al.}(2021){Gaia Collaboration}, {Brown}, {Vallenari}, {Prusti}, {de Bruijne}, {Babusiaux}, {Biermann}, {Creevey}, {Evans}, {Eyer}, {Hutton}, {Jansen}, {Jordi}, {Klioner}, {Lammers}, {Lindegren}, {Luri}, {Mignard}, {Panem}, {Pourbaix}, {Randich}, {Sartoretti}, {Soubiran}, {Walton}, {Arenou}, {Bailer-Jones}, {Bastian}, {Cropper}, {Drimmel}, {Katz}, {Lattanzi}, {van Leeuwen}, {Bakker}, {Cacciari}, {Casta{\~n}eda}, {De Angeli}, {Ducourant}, {Fabricius}, {Fouesneau}, {Fr{\'e}mat}, {Guerra}, {Guerrier}, {Guiraud}, {Jean-Antoine Piccolo}, {Masana}, {Messineo}, {Mowlavi}, {Nicolas}, {Nienartowicz}, {Pailler}, {Panuzzo}, {Riclet}, {Roux}, {Seabroke}, {Sordo}, {Tanga}, {Th{\'e}venin}, {Gracia-Abril}, {Portell}, {Teyssier}, {Altmann}, {Andrae}, {Bellas-Velidis}, {Benson}, {Berthier}, {Blomme}, {Brugaletta}, {Burgess}, {Busso}, {Carry}, {Cellino}, {Cheek}, {Clementini}, {Damerdji}, {Davidson}, {Delchambre}, {Dell'Oro}, {Fern{\'a}ndez-Hern{\'a}ndez}, {Galluccio}, {Garc{\'\i}a-Lario},
  {Garcia-Reinaldos}, {Gonz{\'a}lez-N{\'u}{\~n}ez}, {Gosset}, {Haigron}, {Halbwachs}, {Hambly}, {Harrison}, {Hatzidimitriou}, {Heiter}, {Hern{\'a}ndez}, {Hestroffer}, {Hodgkin}, {Holl}, {Jan{\ss}en}, {Jevardat de Fombelle}, {Jordan}, {Krone-Martins}, {Lanzafame}, {L{\"o}ffler}, {Lorca}, {Manteiga}, {Marchal}, {Marrese}, {Moitinho}, {Mora}, {Muinonen}, {Osborne}, {Pancino}, {Pauwels}, {Petit}, {Recio-Blanco}, {Richards}, {Riello}, {Rimoldini}, {Robin}, {Roegiers}, {Rybizki}, {Sarro}, {Siopis}, {Smith}, {Sozzetti}, {Ulla}, {Utrilla}, {van Leeuwen}, {van Reeven}, {Abbas}, {Abreu Aramburu}, {Accart}, {Aerts}, {Aguado}, {Ajaj}, {Altavilla}, {{\'A}lvarez}, {{\'A}lvarez Cid-Fuentes}, {Alves}, {Anderson}, {Anglada Varela}, {Antoja}, {Audard}, {Baines}, {Baker}, {Balaguer-N{\'u}{\~n}ez}, {Balbinot}, {Balog}, {Barache}, {Barbato}, {Barros}, {Barstow}, {Bartolom{\'e}}, {Bassilana}, {Bauchet}, {Baudesson-Stella}, {Becciani}, {Bellazzini}, {Bernet}, {Bertone}, {Bianchi}, {Blanco-Cuaresma}, {Boch}, {Bombrun}, {Bossini},
  {Bouquillon}, {Bragaglia}, {Bramante}, {Breedt}, {Bressan}, {Brouillet}, {Bucciarelli}, {Burlacu}, {Busonero}, {Butkevich}, {Buzzi}, {Caffau}, {Cancelliere}, {C{\'a}novas}, {Cantat-Gaudin}, {Carballo}, {Carlucci}, {Carnerero}, {Carrasco}, {Casamiquela}, {Castellani}, {Castro-Ginard}, {Castro Sampol}, {Chaoul}, {Charlot}, {Chemin}, {Chiavassa}, {Cioni}, {Comoretto}, {Cooper}, {Cornez}, {Cowell}, {Crifo}, {Crosta}, {Crowley}, {Dafonte}, {Dapergolas}, {David}, {David}, {de Laverny}, {De Luise}, {De March}, {De Ridder}, {de Souza}, {de Teodoro}, {de Torres}, {del Peloso}, {del Pozo}, {Delbo}, {Delgado}, {Delgado}, {Delisle}, {Di Matteo}, {Diakite}, {Diener}, {Distefano}, {Dolding}, {Eappachen}, {Edvardsson}, {Enke}, {Esquej}, {Fabre}, {Fabrizio}, {Faigler}, {Fedorets}, {Fernique}, {Fienga}, {Figueras}, {Fouron}, {Fragkoudi}, {Fraile}, {Franke}, {Gai}, {Garabato}, {Garcia-Gutierrez}, {Garc{\'\i}a-Torres}, {Garofalo}, {Gavras}, {Gerlach}, {Geyer}, {Giacobbe}, {Gilmore}, {Girona}, {Giuffrida}, {Gomel}, {Gomez},
  {Gonzalez-Santamaria}, {Gonz{\'a}lez-Vidal}, {Granvik}, {Guti{\'e}rrez-S{\'a}nchez}, {Guy}, {Hauser}, {Haywood}, {Helmi}, {Hidalgo}, {Hilger}, {H{\l}adczuk}, {Hobbs}, {Holland}, {Huckle}, {Jasniewicz}, {Jonker}, {Juaristi Campillo}, {Julbe}, {Karbevska}, {Kervella}, {Khanna}, {Kochoska}, {Kontizas}, {Kordopatis}, {Korn}, {Kostrzewa-Rutkowska}, {Kruszy{\'n}ska}, {Lambert}, {Lanza}, {Lasne}, {Le Campion}, {Le Fustec}, {Lebreton}, {Lebzelter}, {Leccia}, {Leclerc}, {Lecoeur-Taibi}, {Liao}, {Licata}, {Lindstr{\o}m}, {Lister}, {Livanou}, {Lobel}, {Madrero Pardo}, {Managau}, {Mann}, {Marchant}, {Marconi}, {Marcos Santos}, {Marinoni}, {Marocco}, {Marshall}, {Martin Polo}, {Mart{\'\i}n-Fleitas}, {Masip}, {Massari}, {Mastrobuono-Battisti}, {Mazeh}, {McMillan}, {Messina}, {Michalik}, {Millar}, {Mints}, {Molina}, {Molinaro}, {Moln{\'a}r}, {Montegriffo}, {Mor}, {Morbidelli}, {Morel}, {Morris}, {Mulone}, {Munoz}, {Muraveva}, {Murphy}, {Musella}, {Noval}, {Ord{\'e}novic}, {Orr{\`u}}, {Osinde}, {Pagani}, {Pagano},
  {Palaversa}, {Palicio}, {Panahi}, {Pawlak}, {Pe{\~n}alosa Esteller}, {Penttil{\"a}}, {Piersimoni}, {Pineau}, {Plachy}, {Plum}, {Poggio}, {Poretti}, {Poujoulet}, {Pr{\v{s}}a}, {Pulone}, {Racero}, {Ragaini}, {Rainer}, {Raiteri}, {Rambaux}, {Ramos}, {Ramos-Lerate}, {Re Fiorentin}, {Regibo}, {Reyl{\'e}}, {Ripepi}, {Riva}, {Rixon}, {Robichon}, {Robin}, {Roelens}, {Rohrbasser}, {Romero-G{\'o}mez}, {Rowell}, {Royer}, {Rybicki}, {Sadowski}, {Sagrist{\`a} Sell{\'e}s}, {Sahlmann}, {Salgado}, {Salguero}, {Samaras}, {Sanchez Gimenez}, {Sanna}, {Santove{\~n}a}, {Sarasso}, {Schultheis}, {Sciacca}, {Segol}, {Segovia}, {S{\'e}gransan}, {Semeux}, {Shahaf}, {Siddiqui}, {Siebert}, {Siltala}, {Slezak}, {Smart}, {Solano}, {Solitro}, {Souami}, {Souchay}, {Spagna}, {Spoto}, {Steele}, {Steidelm{\"u}ller}, {Stephenson}, {S{\"u}veges}, {Szabados}, {Szegedi-Elek}, {Taris}, {Tauran}, {Taylor}, {Teixeira}, {Thuillot}, {Tonello}, {Torra}, {Torra}, {Turon}, {Unger}, {Vaillant}, {van Dillen}, {Vanel}, {Vecchiato}, {Viala}, {Vicente},
  {Voutsinas}, {Weiler}, {Wevers}, {Wyrzykowski}, {Yoldas}, {Yvard}, {Zhao}, {Zorec}, {Zucker}, {Zurbach}, \& {Zwitter}}]{Gaia2021A&A}
{Gaia Collaboration}, {Brown}, A.~G.~A., {Vallenari}, A., {et~al.} 2021, Astronomy \& Astrophysics, 649, A1

\bibitem[{{Gentile} {et~al.}(2004){Gentile}, {Salucci}, {Klein}, {Vergani}, \& {Kalberla}}]{Gentile2004MNRAS}
{Gentile}, G., {Salucci}, P., {Klein}, U., {Vergani}, D., \& {Kalberla}, P. 2004, \mnras, 351, 903

\bibitem[{{Guo} {et~al.}(2024){Guo}, {Wei}, {Luo}, {Lin}, {Li}, {Cai}, {He}, {Wang}, \& {Yang}}]{Guo2024arXiv241102161G}
{Guo}, Q., {Wei}, L., {Luo}, W., {et~al.} 2024, arXiv e-prints, arXiv:2411.02161

\bibitem[{{Hikage} {et~al.}(2019){Hikage}, {Oguri}, {Hamana}, {More}, {Mandelbaum}, {Takada}, {K{\"o}hlinger}, {Miyatake}, {Nishizawa}, {Aihara}, {Armstrong}, {Bosch}, {Coupon}, {Ducout}, {Ho}, {Hsieh}, {Komiyama}, {Lanusse}, {Leauthaud}, {Lupton}, {Medezinski}, {Mineo}, {Miyama}, {Miyazaki}, {Murata}, {Murayama}, {Shirasaki}, {Sif{\'o}n}, {Simet}, {Speagle}, {Spergel}, {Strauss}, {Sugiyama}, {Tanaka}, {Utsumi}, {Wang}, \& {Yamada}}]{Hikage_2019PASJ}
{Hikage}, C., {Oguri}, M., {Hamana}, T., {et~al.} 2019, Publications of the Astronomical Society of Japan, 71, 43

\bibitem[{{Ivezi{\'c}} {et~al.}(2019){Ivezi{\'c}}, {Kahn}, {Tyson}, {Abel}, {Acosta}, {Allsman}, {Alonso}, {AlSayyad}, {Anderson}, {Andrew}, {Angel}, {Angeli}, {Ansari}, {Antilogus}, {Araujo}, {Armstrong}, {Arndt}, {Astier}, {Aubourg}, {Auza}, {Axelrod}, {Bard}, {Barr}, {Barrau}, {Bartlett}, {Bauer}, {Bauman}, {Baumont}, {Bechtol}, {Bechtol}, {Becker}, {Becla}, {Beldica}, {Bellavia}, {Bianco}, {Biswas}, {Blanc}, {Blazek}, {Blandford}, {Bloom}, {Bogart}, {Bond}, {Booth}, {Borgland}, {Borne}, {Bosch}, {Boutigny}, {Brackett}, {Bradshaw}, {Brandt}, {Brown}, {Bullock}, {Burchat}, {Burke}, {Cagnoli}, {Calabrese}, {Callahan}, {Callen}, {Carlin}, {Carlson}, {Chandrasekharan}, {Charles-Emerson}, {Chesley}, {Cheu}, {Chiang}, {Chiang}, {Chirino}, {Chow}, {Ciardi}, {Claver}, {Cohen-Tanugi}, {Cockrum}, {Coles}, {Connolly}, {Cook}, {Cooray}, {Covey}, {Cribbs}, {Cui}, {Cutri}, {Daly}, {Daniel}, {Daruich}, {Daubard}, {Daues}, {Dawson}, {Delgado}, {Dellapenna}, {de Peyster}, {de Val-Borro}, {Digel}, {Doherty}, {Dubois},
  {Dubois-Felsmann}, {Durech}, {Economou}, {Eifler}, {Eracleous}, {Emmons}, {Fausti Neto}, {Ferguson}, {Figueroa}, {Fisher-Levine}, {Focke}, {Foss}, {Frank}, {Freemon}, {Gangler}, {Gawiser}, {Geary}, {Gee}, {Geha}, {Gessner}, {Gibson}, {Gilmore}, {Glanzman}, {Glick}, {Goldina}, {Goldstein}, {Goodenow}, {Graham}, {Gressler}, {Gris}, {Guy}, {Guyonnet}, {Haller}, {Harris}, {Hascall}, {Haupt}, {Hernandez}, {Herrmann}, {Hileman}, {Hoblitt}, {Hodgson}, {Hogan}, {Howard}, {Huang}, {Huffer}, {Ingraham}, {Innes}, {Jacoby}, {Jain}, {Jammes}, {Jee}, {Jenness}, {Jernigan}, {Jevremovi{\'c}}, {Johns}, {Johnson}, {Johnson}, {Jones}, {Juramy-Gilles}, {Juri{\'c}}, {Kalirai}, {Kallivayalil}, {Kalmbach}, {Kantor}, {Karst}, {Kasliwal}, {Kelly}, {Kessler}, {Kinnison}, {Kirkby}, {Knox}, {Kotov}, {Krabbendam}, {Krughoff}, {Kub{\'a}nek}, {Kuczewski}, {Kulkarni}, {Ku}, {Kurita}, {Lage}, {Lambert}, {Lange}, {Langton}, {Le Guillou}, {Levine}, {Liang}, {Lim}, {Lintott}, {Long}, {Lopez}, {Lotz}, {Lupton}, {Lust}, {MacArthur}, {Mahabal},
  {Mandelbaum}, {Markiewicz}, {Marsh}, {Marshall}, {Marshall}, {May}, {McKercher}, {McQueen}, {Meyers}, {Migliore}, {Miller}, {Mills}, {Miraval}, {Moeyens}, {Moolekamp}, {Monet}, {Moniez}, {Monkewitz}, {Montgomery}, {Morrison}, {Mueller}, {Muller}, {Mu{\~n}oz Arancibia}, {Neill}, {Newbry}, {Nief}, {Nomerotski}, {Nordby}, {O'Connor}, {Oliver}, {Olivier}, {Olsen}, {O'Mullane}, {Ortiz}, {Osier}, {Owen}, {Pain}, {Palecek}, {Parejko}, {Parsons}, {Pease}, {Peterson}, {Peterson}, {Petravick}, {Libby Petrick}, {Petry}, {Pierfederici}, {Pietrowicz}, {Pike}, {Pinto}, {Plante}, {Plate}, {Plutchak}, {Price}, {Prouza}, {Radeka}, {Rajagopal}, {Rasmussen}, {Regnault}, {Reil}, {Reiss}, {Reuter}, {Ridgway}, {Riot}, {Ritz}, {Robinson}, {Roby}, {Roodman}, {Rosing}, {Roucelle}, {Rumore}, {Russo}, {Saha}, {Sassolas}, {Schalk}, {Schellart}, {Schindler}, {Schmidt}, {Schneider}, {Schneider}, {Schoening}, {Schumacher}, {Schwamb}, {Sebag}, {Selvy}, {Sembroski}, {Seppala}, {Serio}, {Serrano}, {Shaw}, {Shipsey}, {Sick}, {Silvestri},
  {Slater}, {Smith}, {Smith}, {Sobhani}, {Soldahl}, {Storrie-Lombardi}, {Stover}, {Strauss}, {Street}, {Stubbs}, {Sullivan}, {Sweeney}, {Swinbank}, {Szalay}, {Takacs}, {Tether}, {Thaler}, {Thayer}, {Thomas}, {Thornton}, {Thukral}, {Tice}, {Trilling}, {Turri}, {Van Berg}, {Vanden Berk}, {Vetter}, {Virieux}, {Vucina}, {Wahl}, {Walkowicz}, {Walsh}, {Walter}, {Wang}, {Wang}, {Warner}, {Wiecha}, {Willman}, {Winters}, {Wittman}, {Wolff}, {Wood-Vasey}, {Wu}, {Xin}, {Yoachim}, \& {Zhan}}]{Ivezic2019ApJ...873..111I}
{Ivezi{\'c}}, {\v{Z}}., {Kahn}, S.~M., {Tyson}, J.~A., {et~al.} 2019, The Astrophysical Journal, 873, 111

\bibitem[{{Jiang} {et~al.}(2023){Jiang}, {Benson}, {Hopkins}, {Slone}, {Lisanti}, {Kaplinghat}, {Peter}, {Zeng}, {Du}, {Yang}, \& {Shen}}]{JiangF2023MNRAS.521.4630J}
{Jiang}, F., {Benson}, A., {Hopkins}, P.~F., {et~al.} 2023, \mnras, 521, 4630

\bibitem[{{Karukes} \& {Salucci}(2017)}]{Salucci2017MNRAS}
{Karukes}, E.~V. \& {Salucci}, P. 2017, PublicMonthly Notices of the Royal Astronomical Societyations of the, 465, 4703

\bibitem[{{Kendall} \& {Easther}(2020)}]{easther2020PASA}
{Kendall}, E. \& {Easther}, R. 2020, Publications of the Astronomical Society of Australia, 37, e009

\bibitem[{Klypin {et~al.}(2002)Klypin, Zhao, \& Somerville}]{Klypin_2002}
Klypin, A., Zhao, H., \& Somerville, R.~S. 2002, The Astrophysical Journal, 573, 597

\bibitem[{Kroupa(2001)}]{Kroupa2000MNRAS}
Kroupa, P. 2001, Monthly Notices of the Royal Astronomical Society, 322, 231

\bibitem[{Leauthaud {et~al.}(2017)Leauthaud, Saito, Hilbert, Barreira, More, White, Alam, Behroozi, Bundy, Coupon, Erben, Heymans, Hildebrandt, Mandelbaum, Miller, Moraes, Pereira, Rodr{\'{\i} }guez-Torres, Schmidt, Shan, Viel, \& Villaescusa-Navarro}]{Leauthaud_2017MNRAS}
Leauthaud, A., Saito, S., Hilbert, S., {et~al.} 2017, Monthly Notices of the Royal Astronomical Society, 467, 3024

\bibitem[{Mao \& Paczyński(1996)}]{MaoAPJ1996}
Mao, S. \& Paczyński, B. 1996, The Astrophysical Journal, 473, 57

\bibitem[{Mao {et~al.}(2013)Mao, Strigari, Wechsler, Wu, \& Hahn}]{MaoAPJ2013}
Mao, Y.-Y., Strigari, L.~E., Wechsler, R.~H., Wu, H.-Y., \& Hahn, O. 2013, The Astrophysical Journal, 764, 35

\bibitem[{Mr{\'{o}}z {et~al.}(2017)Mr{\'{o}}z, Udalski, Skowron, Poleski, Koz{\l}owski, Szyma{\'{n}}ski, Soszy{\'{n}}ski, Wyrzykowski, Pietrukowicz, Ulaczyk, Skowron, \& Pawlak}]{Mroz2017Nature}
Mr{\'{o}}z, P., Udalski, A., Skowron, J., {et~al.} 2017, Nature, 548, 183

\bibitem[{{Mr{\'o}z} {et~al.}(2020){Mr{\'o}z}, {Udalski}, {Szyma{\'n}ski}, {Soszy{\'n}ski}, {Pietrukowicz}, {Koz{\l}owski}, {Skowron}, {Poleski}, {Ulaczyk}, {Gromadzki}, {Rybicki}, {Iwanek}, \& {Wrona}}]{Mroz2020ApJS}
{Mr{\'o}z}, P., {Udalski}, A., {Szyma{\'n}ski}, M.~K., {et~al.} 2020, The Astrophysical Journals, 249, 16

\bibitem[{Mróz {et~al.}(2019)Mróz, Udalski, Skowron, Szymański, Soszyński, Łukasz Wyrzykowski, Pietrukowicz, Kozłowski, Poleski, Ulaczyk, Rybicki, \& Iwanek}]{Mroz2019APJS}
Mróz, P., Udalski, A., Skowron, J., {et~al.} 2019, The Astrophysical Journal Supplement Series, 244, 29

\bibitem[{{Navarro} {et~al.}(1997){Navarro}, {Frenk}, \& {White}}]{NFW_1997ApJ}
{Navarro}, J.~F., {Frenk}, C.~S., \& {White}, S. D.~M. 1997, The Astrophysical Journal, 490, 493

\bibitem[{{Nesti} \& {Salucci}(2013)}]{Nesti2013JCAP}
{Nesti}, F. \& {Salucci}, P. 2013, \jcap, 2013, 016

\bibitem[{Niikura {et~al.}(2019)Niikura, Takada, Yokoyama, Sumi, \& Masaki}]{Niikura2019PRD}
Niikura, H., Takada, M., Yokoyama, S., Sumi, T., \& Masaki, S. 2019, Physical Review D, 99

\bibitem[{{Novikov} {et~al.}(1979){Novikov}, {Polnarev}, {Starobinskii}, \& {Zeldovich}}]{Starobinskii1979A&A}
{Novikov}, I.~D., {Polnarev}, A.~G., {Starobinskii}, A.~A., \& {Zeldovich}, I.~B. 1979, Astronomy \& Astrophysics, 80, 104

\bibitem[{{Planck Collaboration} {et~al.}(2020){Planck Collaboration}, {Aghanim}, {Akrami}, {Ashdown}, {Aumont}, {Baccigalupi}, {Ballardini}, {Banday}, {Barreiro}, {Bartolo}, {Basak}, {Battye}, {Benabed}, {Bernard}, {Bersanelli}, {Bielewicz}, {Bock}, {Bond}, {Borrill}, {Bouchet}, {Boulanger}, {Bucher}, {Burigana}, {Butler}, {Calabrese}, {Cardoso}, {Carron}, {Challinor}, {Chiang}, {Chluba}, {Colombo}, {Combet}, {Contreras}, {Crill}, {Cuttaia}, {de Bernardis}, {de Zotti}, {Delabrouille}, {Delouis}, {Di Valentino}, {Diego}, {Dor{\'e}}, {Douspis}, {Ducout}, {Dupac}, {Dusini}, {Efstathiou}, {Elsner}, {En{\ss}lin}, {Eriksen}, {Fantaye}, {Farhang}, {Fergusson}, {Fernandez-Cobos}, {Finelli}, {Forastieri}, {Frailis}, {Fraisse}, {Franceschi}, {Frolov}, {Galeotta}, {Galli}, {Ganga}, {G{\'e}nova-Santos}, {Gerbino}, {Ghosh}, {Gonz{\'a}lez-Nuevo}, {G{\'o}rski}, {Gratton}, {Gruppuso}, {Gudmundsson}, {Hamann}, {Handley}, {Hansen}, {Herranz}, {Hildebrandt}, {Hivon}, {Huang}, {Jaffe}, {Jones}, {Karakci}, {Keih{\"a}nen},
  {Keskitalo}, {Kiiveri}, {Kim}, {Kisner}, {Knox}, {Krachmalnicoff}, {Kunz}, {Kurki-Suonio}, {Lagache}, {Lamarre}, {Lasenby}, {Lattanzi}, {Lawrence}, {Le Jeune}, {Lemos}, {Lesgourgues}, {Levrier}, {Lewis}, {Liguori}, {Lilje}, {Lilley}, {Lindholm}, {L{\'o}pez-Caniego}, {Lubin}, {Ma}, {Mac{\'\i}as-P{\'e}rez}, {Maggio}, {Maino}, {Mandolesi}, {Mangilli}, {Marcos-Caballero}, {Maris}, {Martin}, {Martinelli}, {Mart{\'\i}nez-Gonz{\'a}lez}, {Matarrese}, {Mauri}, {McEwen}, {Meinhold}, {Melchiorri}, {Mennella}, {Migliaccio}, {Millea}, {Mitra}, {Miville-Desch{\^e}nes}, {Molinari}, {Montier}, {Morgante}, {Moss}, {Natoli}, {N{\o}rgaard-Nielsen}, {Pagano}, {Paoletti}, {Partridge}, {Patanchon}, {Peiris}, {Perrotta}, {Pettorino}, {Piacentini}, {Polastri}, {Polenta}, {Puget}, {Rachen}, {Reinecke}, {Remazeilles}, {Renzi}, {Rocha}, {Rosset}, {Roudier}, {Rubi{\~n}o-Mart{\'\i}n}, {Ruiz-Granados}, {Salvati}, {Sandri}, {Savelainen}, {Scott}, {Shellard}, {Sirignano}, {Sirri}, {Spencer}, {Sunyaev}, {Suur-Uski}, {Tauber}, {Tavagnacco},
  {Tenti}, {Toffolatti}, {Tomasi}, {Trombetti}, {Valenziano}, {Valiviita}, {Van Tent}, {Vibert}, {Vielva}, {Villa}, {Vittorio}, {Wandelt}, {Wehus}, {White}, {White}, {Zacchei}, \& {Zonca}}]{Planck2020A&A}
{Planck Collaboration}, {Aghanim}, N., {Akrami}, Y., {et~al.} 2020, Astronomy \& Astrophysics, 641, A6

\bibitem[{Riess {et~al.}(1998)Riess, Nugent, Filippenko, Kirshner, \& Perlmutter}]{Riess1998ApJ}
Riess, A.~G., Nugent, P., Filippenko, A.~V., Kirshner, R.~P., \& Perlmutter, S. 1998, The Astrophysical Journal, 504, 935

\bibitem[{Rocha {et~al.}(2013{\natexlab{a}})Rocha, Peter, Bullock, Kaplinghat, Garrison-Kimmel, Oñorbe, \& Moustakas}]{SIDM2013MNRAS}
Rocha, M., Peter, A. H.~G., Bullock, J.~S., {et~al.} 2013{\natexlab{a}}, Monthly Notices of the Royal Astronomical Society, 430, 81

\bibitem[{Rocha {et~al.}(2013{\natexlab{b}})Rocha, Peter, Bullock, Kaplinghat, Garrison-Kimmel, Oñorbe, \& Moustakas}]{Rocha2013MNRAS}
Rocha, M., Peter, A. H.~G., Bullock, J.~S., {et~al.} 2013{\natexlab{b}}, Monthly Notices of the Royal Astronomical Society, 430, 81

\bibitem[{{Salucci}(2019)}]{Salucci2019A&ARv}
{Salucci}, P. 2019, \aapr, 27, 2

\bibitem[{{Salucci} \& {Burkert}(2000)}]{salucci2000ApJ}
{Salucci}, P. \& {Burkert}, A. 2000, \apjl, 537, L9

\bibitem[{{Sasaki} {et~al.}(2018){Sasaki}, {Suyama}, {Tanaka}, \& {Yokoyama}}]{Misao2018CQGra..}
{Sasaki}, M., {Suyama}, T., {Tanaka}, T., \& {Yokoyama}, S. 2018, Classical and Quantum Gravity, 35, 063001

\bibitem[{{Spergel} \& {Steinhardt}(2000)}]{Spergel2000PhRvL}
{Spergel}, D.~N. \& {Steinhardt}, P.~J. 2000, Physical Review Letters, 84, 3760

\bibitem[{{Verde} {et~al.}(2019){Verde}, {Treu}, \& {Riess}}]{Verde2019NatAs}
{Verde}, L., {Treu}, T., \& {Riess}, A.~G. 2019, Nature Astronomy, 3, 891

\bibitem[{{Wang} {et~al.}(2020){Wang}, {Bose}, {Frenk}, {Gao}, {Jenkins}, {Springel}, \& {White}}]{Wang2020Natur}
{Wang}, J., {Bose}, S., {Frenk}, C.~S., {et~al.} 2020, Nature, 585, 39

\bibitem[{{Wang} {et~al.}(2023){Wang}, {Liu}, {Cai}, {Geng}, {Fang}, {He}, {Jiang}, {Jiang}, {Kong}, {Li}, {Li}, {Luo}, {Pan}, {Wu}, {Yang}, {Yu}, {Zheng}, {Zhu}, {Cai}, {Chen}, {Chen}, {Dai}, {Fan}, {Fan}, {Fang}, {He}, {Hu}, {Hu}, {Jin}, {Jiang}, {Li}, {Li}, {Li}, {Liang}, {Lin}, {Liu}, {Liu}, {Liu}, {Liu}, {Liu}, {Lou}, {Qu}, {Sheng}, {Shi}, {Shu}, {Su}, {Sun}, {Wang}, {Wang}, {Wang}, {Wang}, {Wei}, {Wei}, {Xue}, {Yan}, {Yang}, {Yuan}, {Yuan}, {Zhang}, {Zhang}, {Zhao}, \& {Zhao}}]{Wang2023SCPMA}
{Wang}, T., {Liu}, G., {Cai}, Z., {et~al.} 2023, Science China Physics, Mechanics, and Astronomy, 66, 109512

\bibitem[{{Wetzel} {et~al.}(2023){Wetzel}, {Hayward}, {Sanderson}, {Ma}, {Angl{\'e}s-Alc{\'a}zar}, {Feldmann}, {Chan}, {El-Badry}, {Wheeler}, {Garrison-Kimmel}, {Nikakhtar}, {Panithanpaisal}, {Arora}, {Gurvich}, {Samuel}, {Sameie}, {Pandya}, {Hafen}, {Hummels}, {Loebman}, {Boylan-Kolchin}, {Bullock}, {Faucher-Gigu{\`e}re}, {Kere{\v{s}}}, {Quataert}, \& {Hopkins}}]{wetzel2023ApJS}
{Wetzel}, A., {Hayward}, C.~C., {Sanderson}, R.~E., {et~al.} 2023, The Astrophysical Journals, 265, 44

\end{thebibliography}






\appendix
\section{Model dependencies}
In this supplementary material, we will discuss the model dependency of core size and $f_\mathrm{MDMS}$ to show the robustness of our result.

{\it A.1 Velocity distribution --} In the main body of this paper, a Maxwell-Boltzmann distribution for MDMS is assumed. Still, as shown in cosmological simulations, there are other velocity distribution candidates\cite{velocity_distribution_dm}. So here we will discuss how different velocity distributions affect our results.

As shown in Mao et al, 1996\cite{MaoAPJ1996}, for lens with a given velocity $|\vec{v}|$, the asymptotic behaviors of the event rate are given by:
\begin{align}
    \frac{\mathrm{d} \Gamma_{\mathrm{d}}}{\mathrm{d} t_{\mathrm{E}}}
    &\propto \begin{cases}
    (\frac{t_\mathrm{E}}{t_\mathrm{m}})^2 
    & \left(t_\mathrm{E} \ll t_\mathrm{m} \right) 
    \vspace{0.3cm}
    \\
    (\frac{t_\mathrm{E}}{t_\mathrm{m}})^{-4} 
    & \left(t_\mathrm{E} \gg t_\mathrm{m} \right)
    \end{cases}
\end{align}
where the characteristic timescale is $t_\mathrm{m} = \frac{R_\mathrm{E}|_{d_\mathrm{l}= 0.5d_\mathrm{s}}}{|\vec{v}|} $.
If we take the velocity distribution of the lens into account, we would need to integrate over $|\vec{v}|$:
\begin{align}
    \frac{\mathrm{d} \Gamma_{\mathrm{d}}}{\mathrm{d} t_{\mathrm{E}}}
    &\propto \begin{cases}
    t_\mathrm{E}^2\int_0^\infty \frac{v^2\mathrm{d}v}{t_\mathrm{m}^2} f(v)
    & \left(t_\mathrm{E} \ll t_\mathrm{m} \right) 
    \vspace{0.3cm}
    \\
    t_\mathrm{E}^{-4}\int_0^\infty \frac{v^2\mathrm{d}v}{t_\mathrm{m}^{-4}} f(v)
    & \left(t_\mathrm{E} \gg t_\mathrm{m} \right)
    \end{cases}
    \label{vel_dp}
\end{align}
As the velocity only takes value inside a limited interval due to $f(v)$, the integrals hold asymptotically. 

As we can see from Eq. \ref{vel_dp}, for any given velocity distribution, it only results in an overall factor to the event rate for short-timescale and large-timescale events, which is independent of the angular position of the source. As the constraint of core size comes from the angular distribution of event rate, the velocity distribution will have no impact on the prediction of core size when the time scale of events is very short or very large.
That is to say, the impact of velocity distribution on the prediction only comes from those events with $t_\mathrm{E} \sim t_\mathrm{m}$.

To demonstrate the impact of the velocity distribution, here we use the model given in \cite{MaoAPJ2013} for the test:
\begin{align}
    f(|\vec{v}|) 
    & = \begin{cases}
        A \mathrm{exp}(-|\vec{v}|/v_0) (v_\mathrm{esc}-|\vec{v}|^2)^p,
        & 0\leq |\vec{v}| \leq v_\mathrm{esc}\\
        0,
        & \rm otherwise
    \end{cases}
\end{align}
The result is shown in FIG. \ref{fig:rr_vel}. As we can see, even though the new model has a different mean velocity and velocity dispersion, the one-sigma range still overlaps with the previous one. This indicates the robustness of our result.
\begin{figure}[]
\includegraphics[width=0.95\columnwidth]{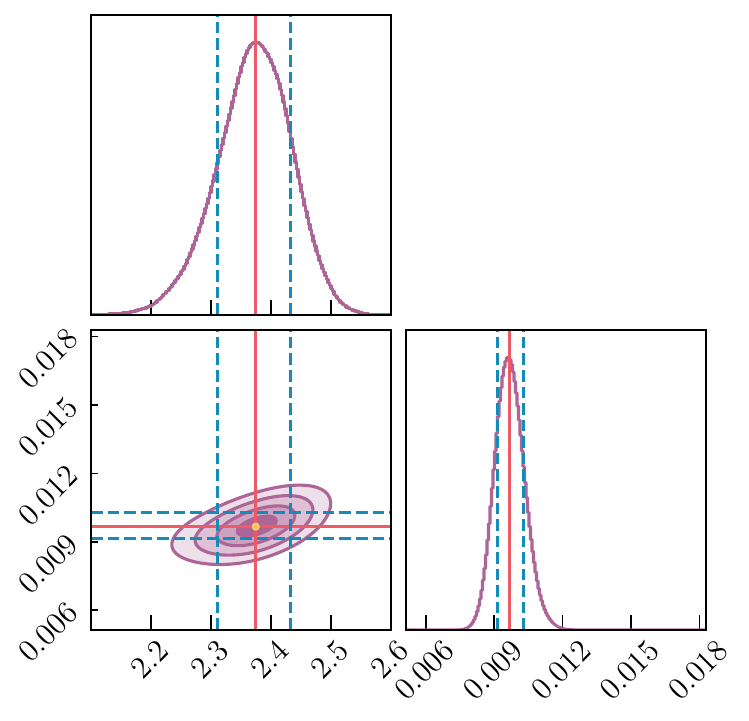}
    \caption{We use the velocity distribution given in \cite{MaoAPJ2013}, with 
    $(v_0/v_\mathrm{esc}, p) = (0.13, 0.78)$ and $M_\mathrm{lens} = 10^{-2.2} M_\odot$. Here the core size is $237^{+34}_{-32}$ ($2.37^{+0.06}_{-0.06}$ in magnitude) and the $f_\mathrm{MDMS} = 0.0096^{+0.0006}_{-0.0005}$. The one-sigma range still overlaps with the Gaussian cases.
    }
\label{fig:rr_vel}
\end{figure}

{\it A.2 Extended lens --} For a point lens, the criterion for a micro-lensing event is:
\begin{equation}
\Delta_\mathrm{sl}<R_\mathrm{E}
\label{eq:cr}
\end{equation}
Here, $\Delta_\mathrm{sl}$ is the distance between the lens and the projection of the source in the lens plane. $R_\mathrm{E}$ is the Einstein radius.

\begin{figure}
\includegraphics[width=0.95\columnwidth]{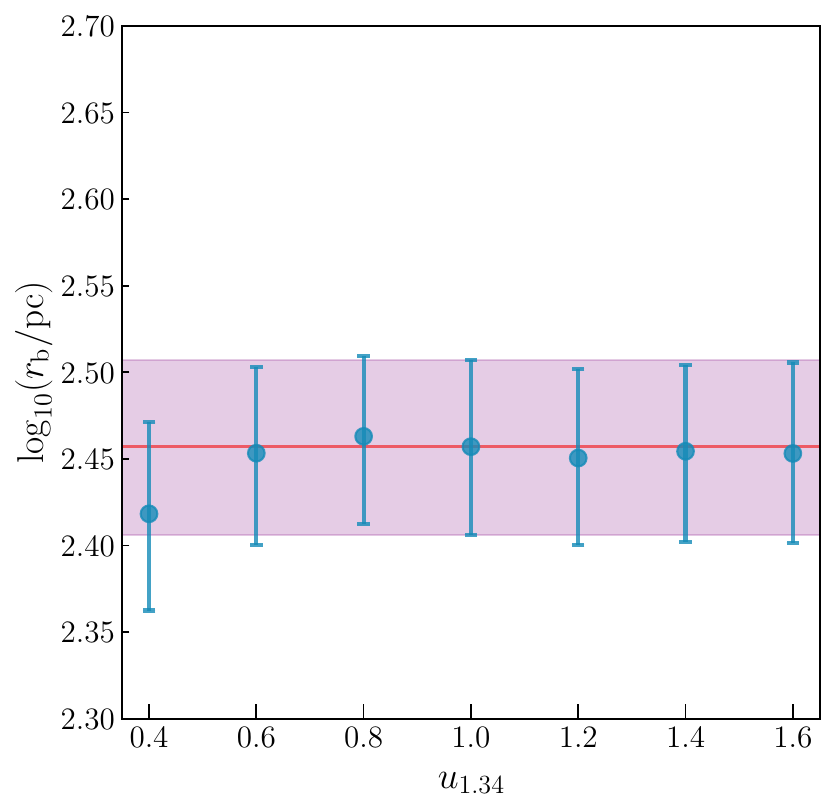}
    \caption{Here we have done tests with $u_{1.34} = 0.4, 0.6, 0.8, 1, 1.2, 1.4, 1.6$ and $M_\mathrm{lens} = 10^{-1.0}M_\odot$. As we can see, for different impact parameters, the results are still within the one sigma range of the fiducial value(i.e. $u_{1.34} = 1.0$.)
    }
\label{fig:Re_test}
\end{figure}

When $\Delta_\mathrm{sl}=R_\mathrm{E}$, the total magnification $\mu_\mathrm{T} = 1.34$. If we keep this criterion, as shown in Croon et al, 2020\cite{CroonPRD2020}, the effect of an extended lens will result in an impact parameter $u_{1.34}$ which serves as a modifying factor to $R_\mathrm{E}$. So, the micro-lensing event rate for extended will be:
\begin{align}
\nonumber
    \frac{\mathrm{d} \Gamma_{\mathrm{d}}}{\mathrm{d} t_{\mathrm{E}}}
    &=\pi
    \int \mathrm{d} (\ln M) \frac{\mathrm{d} n_{\mathrm{lens}}(M)}{\mathrm{d} \ln M}
    \int_{0}^{d_{\mathrm{s}}}\mathrm{d}d_{1}\frac{ \rho_{\mathrm{lens}}\left(d_{1}\right)}{M}\\
    &\int_{-\pi / 2}^{\pi/2}\mathrm{~d\theta}(v'_{\perp})^{4}f_{\mathrm{lens}}\left(v_{\perp}',\theta \right)
\end{align}
Here $v'_\bot =(2u_{1.34}R_{\mathrm{E}} \cos\theta)/(t_\mathrm{E})$.

To test the impact of the possible extended structure of MDMS, we take $u_{1.34} = 0.4, 0.6, 0.8, 1, 1.2, 1.4, 1.6$ ($u_{1.34}$ is the case we showed in the main body of our paper) and rerun the whole analysis. The result is shown in FIG. \ref{fig:Re_test}. For all the impact parameters we tested, the results are all within the one-sigma range of the point lens case(i.e. $u_{1.34} = 1.0$). This indicates that our result is still valid even when considering the extended structure of MDMS.

{\it A.3 Anisotropy of density profile --} \
As discussed in Mr\'{o}z et al. 2020 \cite{Mroz2020ApJS}, the event rate observed in the northern hemisphere is lower than in the southern hemisphere. To examine the impact of this anisotropy, we consider the first-order perturbation of the density profile of MDMS:
\begin{align}
    \rho_\mathrm{aniso}(x,y,z)
    & = 
    \begin{cases}
        (1+\alpha) \rho_\mathrm{iso}(x,y,z)
        & z>0\\
        (1-\alpha) \rho_\mathrm{iso}(x,y,z)
        & z<0
    \end{cases}
    \label{eq:aniso}
\end{align}
Here $\alpha$ is used to demonstrate the first-order anisotropy in the density profile
\begin{figure}
\includegraphics[width=0.95\columnwidth]{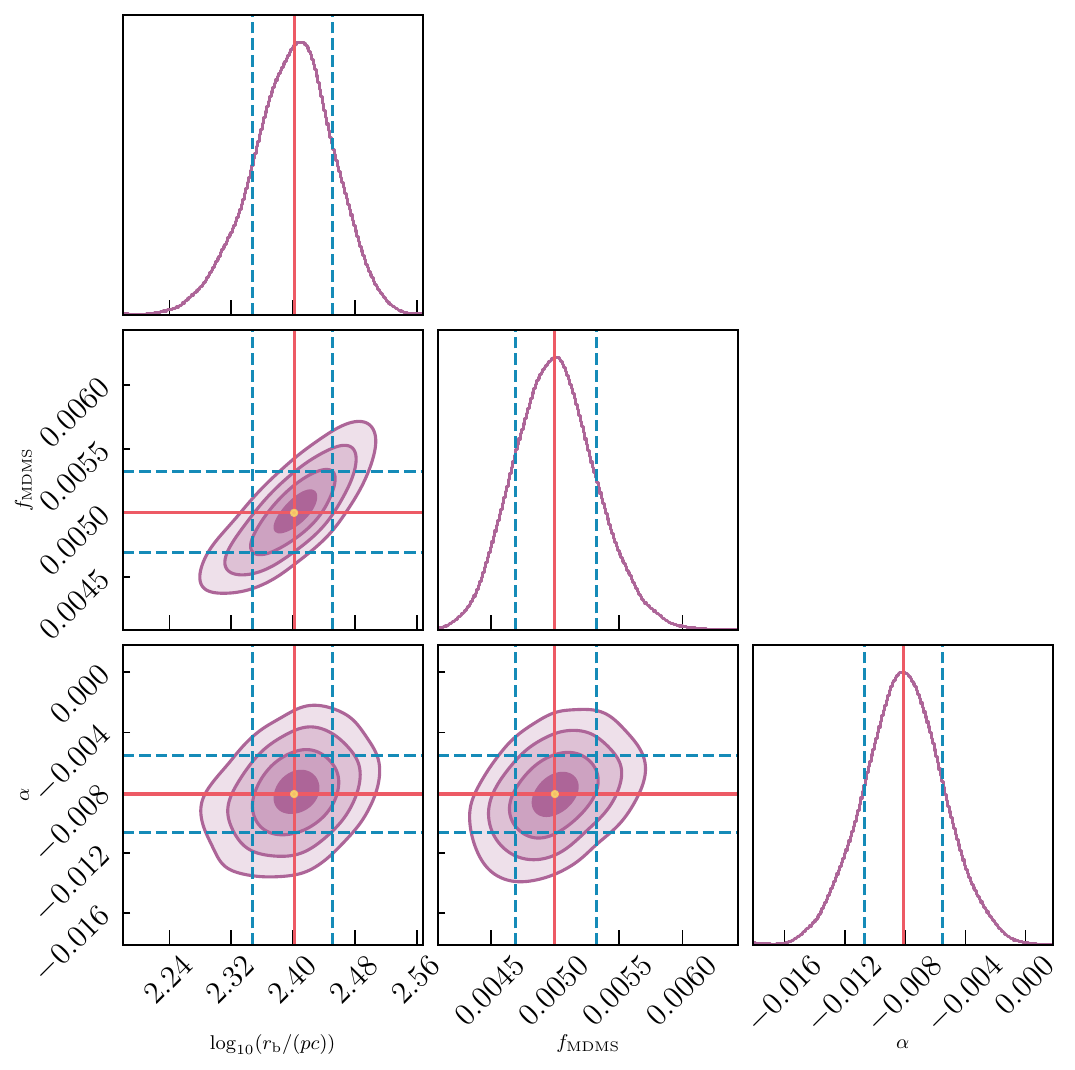}
    \caption{We use $\rho_\mathrm{aniso}$ defined in Eq. \ref{eq:aniso} to rerun the MCMC with $\alpha$ as an additional parameter. Here, $M_\mathrm{lens} = 10^{-2.2} M_\odot$. The core size is $252^{+30}_{-30}$ ($2.40^{+0.05}_{-0.05}$ in magnitude), $f_\mathrm{MDMS} = 0.0050^{+0.0003}_{-0.0003}$, and $\alpha = -0.008^{+0.002}_{-0.002}$. This is still within the one-sigma range shown in the main body of our paper.
    }
\label{fig:rr_ansio}
\end{figure}

With this profile, we rerun the whole analyzing process with $\alpha$ as an additional parameter in the MCMC process. The result is shown in FIG. \ref{fig:rr_ansio}. The result shows the density in the southern hemisphere is a little higher than the northern part ($\alpha = -0.008$), which coincides with the OGLE observation. 
Also, the new core size, $240$ pc, is still within the one-sigma range shown in the main body of our paper. So, to the first-order level, the core size we present here is not significantly affected by the anisotropy.

\end{document}